\documentclass[twocolumn,aps,floatfix,showpacs,prd,longbibliography]{revtex4-1}
\usepackage{graphics, epsfig ,color} 
\usepackage{amsmath,amssymb} 
\usepackage{amsfonts,subfigure}
\usepackage{ulem}
\usepackage{enumerate}
\usepackage[dvipsnames]{xcolor}

\newcommand{\av}[1]{\langle\, #1 \, \rangle}

\newcommand{\bra}[1]{\left[\, #1 \, \right]}
\newcommand{\pare}[1]{\left(\, #1 \, \right)}

\graphicspath{{figures/}}

\begin{document}
\date{\today}
\title{Noise driven broadening of the neural synchronisation transition in stage II retinal waves}
\author{Dora Matzakos-Karvouniari$^{1,3}$, Bruno Cessac$^{1}$ and L. Gil$^{2}$ }
\affiliation{$^1$ Universit\'e C\^ote d'Azur, Inria, Biovision team, France\\
$^2$ Universit\'e C\^ote d'Azur, Institut de Physique de Nice (INPHYNI), France\\
$^3$ Universit\'e C\^ote d'Azur, Laboratoire Jean-Alexandre Dieudonn\'e (LJAD), France}

\begin{abstract}
Based on a biophysical model of retinal Starburst Amacrine Cell (SAC) \cite{karvouniari-gil-etal:19} we analyse here the dynamics of retinal waves, arising during the visual system development. Waves are induced by spontaneous bursting of SACs and their coupling via acetycholine. We show that, despite the acetylcholine coupling intensity has been experimentally observed to change during development \cite{zheng-lee-etal:04}, SACs retinal waves can nevertheless stay in a regime with power law distributions, reminiscent of a critical regime. Thus, this regime occurs on a range of coupling parameters instead of a single point as in usual phase transitions. 
We explain this phenomenon thanks to a coherence-resonance mechanism, where noise is responsible for the broadening of the critical coupling strength range.

\end{abstract}
\maketitle

\section{Introduction}\label{Sec:Intro}

Since the seminal work of Beggs and Plen \cite{beggs-plenz:03} - reporting that neocortical activity in rat slices occur in the form of neural avalanches with power law distributions close to a critical branching  process -
there have been numerous papers suggesting that the brain as a dynamical system fluctuates around a critical point.
Scale-free neural avalanches have been found to occur in a wide range of neural tissues and species \cite{shew-plenz:13}. From a theoretical point of view, it has been suggested \cite{haldeman-beggs:05,shew-plenz:13} that such a scale-free organisation could foster information storage and transfer, improvement of the computational capabilities \cite{bertschinger-natschlager:04}, information transmission \cite{beggs-plenz:03,bertschinger-natschlager:04,shew-yang-etal:11}, sensitivity  to  stimuli  and  enlargement  of  dynamic  range  \cite{kinouchi-copelli:06,shew-yang-etal:09,gautam-hoang-etal:15,girardi-schappo-bortolotto-etal:16}. 

In this spirit  it has been proposed by M. Hennig et al, in a paper combining experiments and modelling, that such a critical regime could also take place in the visual system, at the early stages of its development \cite{hennig-adams-etal:09}. Investigating the dynamics of stage II retinal waves, a mechanism participating in the development of the visual system of mammals \cite{wong-meister-etal:93, firth-wang-etal:05, sernagor-hennig:12, ford-feller:12} (see section \ref{Sec:Biophys} for a detailed description), they show that the network of neurons (here Starburst Amacrine Cells - SACs) is capable of operating at a transition point between purely local and global functional connectedness, corresponding to a percolation phase transition, where waves of activity - often referred to as "avalanches"- are distributed according to power laws (see Fig. 4 of \cite{hennig-adams-etal:09}). They interpret this regime as an indication that early spontaneous activity in the developing retina is regulated according to the following principle; maximize randomness and variability in the resulting activity patterns. This remark is in complete agreement
with the idea of dynamic range maximization \cite{ribeiro-copelli:08} and could be of central importance for our understanding of the visual system. In fact, it suggests that, during its formation, the visual system could be driven by spontaneous events, namely the retinal waves, exhibiting the characteristics of a second order phase transition.

This captivating idea raises nevertheless the following important issue. The spontaneous stage II retinal waves are mediated by a \textit{transient} network of SACs, connected through excitatory cholinergic connections \cite{zheng-lee-etal:06}, which are formed only during a developmental window up to their complete disappearance. Especially, in \cite{zheng-lee-etal:04}, Zheng et al have shown that the intensity of the acetylcholine coupling is \textit{monotonously decreasing} with time. However, the notion of a critical state corresponds to a \textit{point} in the control parameter space of a system. Assuming here the cholinergic coupling as a control parameter, the above narrative is incompatible with the framework of criticality. In contrast, Zheng et al observations would suggest that criticality, if any, (a) either lasts for a short moment during development where the coupling parameter is \textit{right} at a critical value, ruining any hope of robustness of the phenomenon, or (b) occurs within an \textit{interval} of this coupling parameter. 

Two theoretical arguments could solve this apparent incompatibility: (i) There is a hidden mechanism adapting to the coupling parameter variations so as to maintain the retina in a critical state, in a mechanism reminiscent to self-organized criticality (SOC) e.g. the scenario called "Mapping Self-Organized Criticality onto Criticality" introduced by D. Sornette  and co-workers in \cite{sornette-johansen-etal:97}. In the past, this line of thought has collected a
broad consensus and resulted in numerous interesting results \cite{sornette-johansen-etal:97,levina-herrmann-etal:07,levina-herrmann-etal:09,di-santo-villegas-etal:18,rubinov-sporns-etal:11,shew-clawson-etal:15}. 
For example, in \cite{levina-herrmann-etal:07} and more recently in \cite{di-santo-villegas-etal:18}, dynamical synapses are shown to be responsible for self-organized criticality. In \cite{kossio-goedeke-etal:18}, the feedback of the control parameter to the order parameter is carried out through the dynamics of the synaptic tree radius. (ii) Retinal waves dynamics indeed displays power law distributions on an \textit{interval} of the coupling parameter, without the need of adding an extra mechanism.

In this paper we report on the action of noise fluctuations onto the retinal waves dynamics and give strong arguments in favor of hypothesis (ii). In the past, noise has already be shown to lead to unexpected behavior in excitable systems. In 1997, studying the dynamics of a single excitable FitzHugh-Nagumo neuron under external noise driving, Pikovsky and Kurths  \cite{pikovsky-kurths:97} reported the existence of a {\it coherence resonance} mechanism, characterised by the existence of a noise amplitude for which the self-correlation characteristic time displays a maximum.  Moreover, for this noise amplitude, the signal to noise ratio displays a minimum. Later, on neural networks cell cultures and in computer simulations, the existence of an optimal level of noise  for which the regularity of the synchronized neural interburst interval becomes maximal, has been reported \cite{kim-lee-etal:15}. 

Here, we argue that noise is responsible for the broadening of the region of cholinergic coupling where stage II retinal waves exhibit power laws. Starting from a model we developed in \cite{karvouniari-gil-etal:19} describing the individual bursting behavior of Starburst Amacrine Cells (SACs) and reproducing numerous experimental observations, we study the properties of stage II retinal waves and their dependence upon the cholinergic coupling. The existence of a phase transition between asynchronized and synchronized pulse coupled excitable oscillators has already been well established in the literature \cite{mirollo-strogatz:90,kuramoto:91,corral-perez-etal:95}. We do observe it as well as the usual power law behaviours are commonly observed at the threshold of phase transition. But here, because of the system's closeness to a saddle node bifurcation point, thoroughly analysed in \cite{karvouniari-gil-etal:19}, retinal waves can be initiated by noise by a mechanism detailed below. This leads to a surprising consequence: in the absence of any SOC-like mechanism we numerically show the existence of a whole interval of cholinergic coupling where retinal waves exhibit power laws. The width of this interval depends on noise and there is an optimal noise level where this interval has a maximal width. We call this effect {\it noise driven broadening of the neural synchronisation transition}.

The paper is organised as follows. In section \ref{Sec:Model} we briefly outline the conductance-based dynamical model previously used in \cite{karvouniari-gil-etal:19} to describe the individual SAC dynamics and quickly remind the main associated results.  We then introduce the synaptic cholinergic coupling, discuss the complexity of the ensuing model and the various difficulties associated with its numerical simulation and conclude with the validity limits of our numerical investigation. Section \ref{Sec:SynchroPhase} deals with the neural synchronisation phase transition. Numerical evidence of such a phase transition is given from the point of view of global firing rate (i.e. the total number of spiking SACs at a given time) and evolution of the shape of the avalanches size probability density function (pdf). In section \ref{Sec:TravellingWavesNucleation} we discuss these results from the point of view of travelling waves \cite{bressloff-coombes:98} and give a theoretical interpretation leading to the correct prediction of a range of critical cholinergic coupling strengths.  Finally, section \ref{Sec:RoleNoise} is devoted to the role played by the noise fluctuations in the broadening of the synchronisation transition region. In the conclusion, we discuss the links of our result with self-organized criticality and similar mechanisms that could explain the broadening of the critical range. We also briefly discuss the geometry of waves dynamics and their potential links to visual system development.

\section{Stage II retinal waves model}\label{Sec:Model}

\subsection{The biophysics of retinal waves}\label{Sec:Biophys}
Retinal waves are bursts of activity occurring spontaneously in the developing retina of vertebrate species, contributing to the shaping of the visual system organization \cite{wong-meister-etal:93, firth-wang-etal:05, sernagor-hennig:12, ford-feller:12}. They are characterized by localized groups of neurons becoming simultaneously active, initiated at random points and propagating at speeds ranging from $100 \mu m/s$ (mouse, \cite{singer-mirotznik-etal:01}, \cite{maccione-hennig-etal:14}) up to $400 \mu m/s$ (chick, \cite{sernagor-eglen-etal:00}), with changing boundaries, dependent on local refractoriness \cite{ford-feller:12,ford-felix-etal:12}. This activity, slowly spreading across the retina, is an inherent property of the retinal network \cite{zheng-lee-etal:06}. 
More precisely, the generation of waves requires three conditions \cite{ gjorgjieva-eglen:11,ford-feller:12}: 
\begin{enumerate}[(C1)]
\item A source of depolarization for wave initiation (\textit{"How do waves start?"}). Given that there is no external input (e.g. from visual stimulation in the early retina), there must be some intrinsic mechanism by which neurons become active;
\item  A network of excitatory interactions for propagation (\textit{"How do waves propagate?"}). Once some neurons become spontaneously active, how do they excite neighboring neurons ?
\item  A source of inhibition that limits the spatial extent of waves and dictates the minimum interval between them (\textit{How do waves stop?}). 
\end{enumerate}
 Wave activity begins in the early development, long before the retina is responsive to light. It emerges due to several biophysical mechanisms which change during development, dividing retinal waves maturation into 3 stages (I, II, III) \cite{sernagor-hennig:12}. Each stage, mostly studied in mammals, is characterized by a certain type of network interaction (condition C2): gap junctions for stage I; cholinergic transmission for stage II; and glutamatergic transmission for stage III. In this work, we focus on stage II. 

During this period, the principal mechanism for transmission is due to the neurotransmitter acetylcholine. The first functional cholinergic connections are formed around birth, firstly at Starbust Amacrine Cells (SACs). The emergence of waves depends on cellular mechanisms studied by \cite{zheng-lee-etal:06} (for the rabbit), where it is found that Starburst Amacrine Cells emit spontaneous intrinsic Calcium bursts (condition C1). When the multiple bursts occuring in a neighbourhood of a neuron are synchronized, they trigger a wave. Stage II waves are therefore spatiotemporal phenomena resulting from the spatial coupling of local bursters (SACs) via acetylcholine (condition C2). In addition, a strong after hyperpolarization current (sAHP)  induces a long refractory period for the recently active neurons, preventing the propagation of a new wave for a period of order few seconds, in the region where a wave has spread  \cite{zheng-lee-etal:06}. Moreover, sAHP plays a prominent role in shaping waves periodicity and boundaries (condition C3). 

\subsection{The model}\label{Sec:ModelDef}

The model we propose is initially inspired by the work of \cite{lansdell-ford-etal:14} and \cite{hennig-adams-etal:09}, with strong variations, justified on biophysical grounds. It accurately reproduces experimental facts in stage II retinal waves and affords their mathematical study with qualitative and quantitative descriptions of points C1, C2, C3 \cite{karvouniari:18,karvouniari-gil-etal:19}. 
It
describes a network of Starbust Amacrine Cells (SAC) distributed on a regular lattice where sites/neurons are labelled with an index $j=1 \dots N$.
The state variables characterizing neuron $j$ are: $V_j(t)$, the membrane potential, $N_j(t)$, the gating variable for fast $K^+$ channels, $R_j(t)$ and $S_j(t)$, the gating variables for slow $Ca^{2+}$-gated $K^+$ channels, $C_j(t)$, the intracellular $Ca^{2+}$ concentration, $A_j(t)$, the extracellular acetylcholine concentration emitted by neuron $j$.  We note $\left\{ j \right\}$ the set of SACs which are in synaptic contact with SAC $j$ ($j {\not \in} \left\{ j \right\}$) and $I_{{A}_{j}}(t)$ the acetylcholine synaptic current seen by cell $j$.
All parameter values are carefully calibrated with respect to biophysics, found in the literature or fitted from experimental curves in \cite{abel-lee-etal:04}, \cite{zheng-lee-etal:04} and \cite{zheng-lee-etal:06}. Parameters values and units are given in appendix \ref{AppendixParameter} \\

The equations ruling the dynamics are: 
\begin{equation}
\left\{
\begin{array}{l}
\begin{array}{lcl}
C_{m} {{dV_{j}(t)}\over{dt}} &= &-g_{L}  \left(V_{j}(t)-V_{L}\right) 
\cr
& & -g_{C} M_{\infty} \bra{V_{j}(t)} \, \pare{V_{j}(t)-V_{C}}
\cr
 & & - g_{K} N_{j}(t) \, \pare{V_{j}(t)-V_{K}}
 \cr
 & &\underbrace{- g_{H} \, R_{j}(t)^4 \, \pare{V_{j}(t)-V_{K}}}_{I_{H_{j}}(t)}
  \cr
 & & \underbrace{- g_{A} \left(V_{j}(t)-V_{A}\right) \displaystyle{\sum_{k \in \left\{ j \right\} } {{A_{k}(t)^{2}}\over{\gamma_{Ach}+A_{k}(t)^2}}}}_{I_{{A}_{j}}(t)}
 \cr 
 & & + \eta \, \xi_{j}(t),
 \end{array}
 \cr \cr
 \tau_{N} {{dN_{j}(t)}\over{dt}}=\Lambda\bra{V_{j}(t)} \pare{N_{\infty}\bra{V_{j}(t)}-N_{j}(t)},
  \cr \cr
 \tau_{R} {{dR_{j}(t)}\over{dt}}=\alpha_{R} S_{j}(t) \left(1-R_{j}(t)\right) -R_{j}(t),
\cr \cr
 \tau_{S} {{dS_{j}(t)}\over{dt}}=\alpha_{S} C^4_{j}(t) \left(1-S_{j}(t)\right) -S_{j}(t),
 \cr \cr
 \tau_{C} {{dC_{j}(t)}\over{dt}}=-{{\alpha_{C}}\over{H_{x}}} C_{j}(t) +C_{0} -\delta_{C} g_{C} M_{\infty}\bra{V_{j}(t)} \, \left(V_{j}(t)-V_{C}\right),
 \cr \cr
 {{dA_{j}(t)}\over{dt}}=-\mu A_{j}(t)+\beta_{Ach} T_{Ach}\bra{V_{j}(t)},
 \end{array}
 \right.
\label{eq:FullModel}
\end{equation}

with
\begin{equation}
\left\{
\begin{array}{l}
M_{\infty}\bra{V}={{1}\over{2}}\left(1+\tanh\left({{V-V_{1}}\over{V_{2}}}\right)\right),
\cr
\Lambda\bra{V}=\cosh\left({{V-V_{3}}\over{2V_{4}}}\right),
\cr
N_{\infty}\bra{V}={{1}\over{2}}\left(1+\tanh\left({{V-V_{3}}\over{V_{4}}}\right)\right),
\end{array}
\right.
\end{equation}

and
\begin{equation}
T_{Ach}\bra{V}={{1}\over{1+exp\left(-K_{Ach}\left(V-V_{0}\right)\right)}}.
\end{equation}

$C_{m}$ is the membrane capacitance; $g_{L}$ and $V_{L}$ are, respectively, the leak conductance and the leak reversal potential; $g_{C} M_{\infty}\bra{V}$ is the voltage dependent conductance  and $V_{C}$ the $Ca^{2+}$ reversal potential of $Ca^{2+}$ ionic channels; $g_{K}$ and $V_{K}$ are the conductance and reversal potential associated with fast voltage-gated $K^{+}$ channels; $g_{H}$ is the maximum SAHP conductance. 

Voltage dynamics is stochastic: $\xi_{j}$  are independent, Gaussian, random variables, that mimic noise in dynamics, with vanishing average and a variance equal to 1. The parameter $\eta$ stands for the noise amplitude.

The term $I_{{A}_{j}}$ corresponds to the acetylcholine current (with nicotinic receptors explaining the power $2$ in the term $A_k^2$).  $g_{A}$ stands for the acetylcholine coupling strength and is the main control parameter of our model. Note that, while $g_{L}$, $g_{C}$, $g_{K}$, $g_{H}$ are cell membrane conductances, $g_{A}$ is a conductance per synapse. At rest, for $V_{j}$ close to $V_{L}$, $T_{Ach}\bra{V_{j}}\simeq 0$ and the acetylcholine production is almost vanishing. On the contrary, when voltage increases during bursting, $V_{j}\simeq 0 \,mV$, the production of Ach sharply increases generating the  excitatory current $I_{{A}_{j}}(t)$. 

Finally, $I_{H_j}$ is a slow, potassium-gated, hyper-polarization current (sAHP) limiting the bursting period and waves propagation. 

\subsection{Dynamics of an isolated SACs}\label{Sec:DynIsolated}

The two key biophysical mechanisms associated with the emergence of spontaneous activity during early retinal development are (i) the fast repetitive bursts of spikes mainly controlled by fast voltage-gated channels and (ii) the existence of the prolonged slow after hyper polarisation (sAHP) modulating fast oscillations, controlled by $Ca^{2+}$-gated $K^+$ channels \cite{zheng-lee-etal:06}. These two effects are reproduced by the model \eqref{eq:FullModel} and the dynamics of isolated SACs ($g_{A}=0$) has been thoroughly analysed in \cite{karvouniari-gil-etal:19}. 
Bifurcations analysis allowed us to mathematically understand the origin and dynamics of bursts, and the role of sAHP. We give here the method and the main conclusions of that paper, useful to understand as well the wave dynamics in the presence of acetylcholine.

\begin{figure}[htbp!]
\centerline{
\includegraphics[width=7cm,height=8cm]{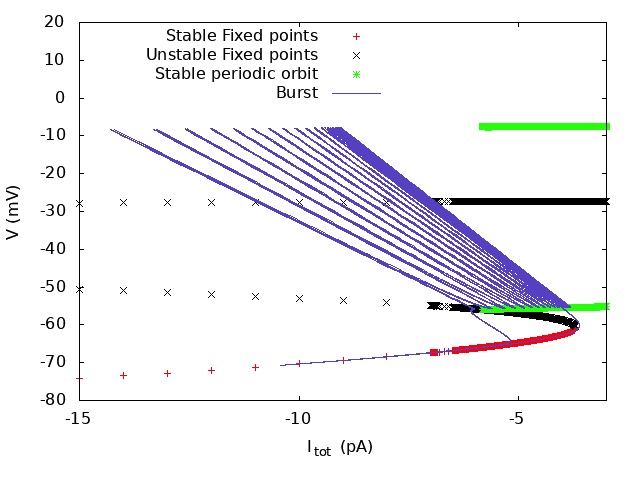}
\hspace{0.3cm}
}
 \caption{\label{Fig:SpontaneousBursting}
  Representation of bursting in the plane $I_{tot}-V$ in relation with the bifurcation diagram of eq.\ref{eq:FastEquations}. In abscissa, the instantaneous current $I_{tot}$ and in ordinate the instantaneous voltage. The merging of the red branch (stable fixed points) with the black branch (unstable fixed points) corresponds to a saddle-node bifurcation. The green line corresponds to the minimum and maximum voltage amplitude of a stable periodic orbit. This orbit collapses with the unstable branch of fixed points via a homoclinic bifurcation. Purple traces correspond to the trajectory of a burst in the two dimensional plane $I_{tot},V$ where $I_{tot}$ depends on $V$ via the sAHP current.
}
\end{figure}

The analysis is based on time scales separation. There are indeed $3$ distinct timescales: (1) fast variables ($\sim 10 ms$) $V_j(t),N_j(t)$; (2) medium scale variable ($\sim 2$ s) $C_j(t),A_j(t)$; (3) slow variables ($\sim 10$ s) $R_j(t)$ and $S_j(t)$. Taking advantage of this huge time scale separation, the fast dynamics of variables $V_j,N_j$ can be rewritten as:
\begin{equation}
\left\{
\begin{array}{l}
\begin{array}{lcl}
C_{m} {{dV_j(t)}\over{dt}} &= &-g_{L}  \left(V_j(t)-V_{L}\right)  \cr
& & - g_{C} M_{\infty}\bra{V_j(t)} \pare{V_j(t)-V_{C}}
 \cr
 & & - g_{K} N_j(t) \pare{V_j(t)-V_{K}} + I_{tot,j}
  \end{array}
 \cr \cr
 \tau_{N} {{dN_j(t)}\over{dt}}=\Lambda\bra{V_j(t)} \pare{N_{\infty}\bra{V_j(t)}-N_j(t)}.
  \cr \cr
  \end{array}
 \right.
\label{eq:FastEquations}
\end{equation}
where $I_{tot,j}=I_{H_{j}}+I_{{A}_{j}}$ is slowly varying compared to the fast scale dynamics. Eq. \eqref{eq:FastEquations} correspond to a Morris-Lecar model \cite{morris-lecar:81} whose bifurcations under the variations of several parameters has been analysed in \cite{karvouniari-gil-etal:19}. Considering $I_{tot,j}$ as such a parameter one can show the following (fig. \ref{Fig:SpontaneousBursting}). There exist two regimes separated by a small transition interval  $[I_{HC},I_{SN}]$. $I_{HC}$ corresponds to an homoclinic bifurcation, $I_{SN}$ to a saddle-node bifurcation. When $I_{tot,j} <I_{HC}$ the neuron is in a rest state with a voltage close to $V_L$. 
For   $I_{tot,j} > I_{SN}$ there is a stable limit cycle corresponding to fast voltage oscillations. In the interval $[I_{HC},I_{SN}]$ a stable fixed point coexists with a stable limit cycle. In this region, when increasing $I_{tot,j}$, the stable fixed point disappears for $I_{tot,j}=I_{SN}$, while,  decreasing $I_{tot,j}$, the limit cycle disappears for $I_{tot,j}=I_{HC}$.

When coupled with the slow sAHP current isolated neurons produce bursts, as illustrated in the figure \ref{Fig:SpontaneousBursting}. Here is the mechanism. Assume first that a SAC spontaneously produces fast oscillations (i.e. dynamics lives on stable limit cycle). As this neuron is in a high voltage state, calcium loads, leading to a raise of sAHP (a decrease of $I_{H_{j}}$ becoming more and more negative) which eventually leads the cell through the homoclinic bifurcation and bursting stops. $I_{H_{j}}$ goes on decreasing for a while, leading to hyperpolarization. Then, because voltage is low, $[Ca^{2+}]$ decreases and $I_{H_{j}}$ increases until the cell returns to its rest state. 

The cell stays in the rest state in the absence of noise. For the parameters values chosen here this rest state is however close to the saddle-node bifurcation $SN$ so that noise will eventually induce the neuron to jump back to the fast oscillating regime. The bifurcation diagram explains therefore the bursting mechanisms and  how noise triggers it (condition C1); it explains as well the role of sAHP in controlling the bursting period (see \cite{karvouniari-gil-etal:19}) (condition C2).\\

Note that, while a low noise intensity $\eta$ has the effect of triggering bursting, a too high intensity of noise can, on the contrary, stop bursting. The mechanism was illustrated in the bifurcation diagram fig.11d of the paper \cite{karvouniari-gil-etal:19}: the noise is so high that the neuron cannot stay on the limit cycle long enough to trigger the calcium and sAHP activation. Therefore, the bursting period depends on $\eta$, \textit{increasing first, then decreasing}. Since, the medium scale phenomena (calcium and acetylcholine production) requires that the SAC stays in a fast oscillation regime long enough, they take place only in a range of value of $\eta$ and are not possible when $\eta$ becomes too large.

\subsection{Dynamics of coupled SACs}\label{Sec:DynCoupled}
We now consider the dynamics of SACs in the presence of acetylcholine coupling ($g_{A}>0$). When a SACs is bursting, in a high voltage state, it releases acetylcholine inducing an excitatory current in the neighbouring SACs, increasing their excitability. When the amount of neurotransmitter is high enough, the neighbouring SACs can be excited enough to start to burst thereby inducing a wave of activity propagating through the lattice. 
Hence the system of equations (\ref{eq:FullModel}) belongs to the well known class of pulse-coupled excitable oscillator which is known to possess travelling wave solution \cite{bressloff-coombes:98}.

\subsection{Numerical simulations}
\label{Sec:NumericalSimulations}

Starburst amacrine cell (SAC) display an almost 2D isotropic synaptic tree with a disk shape. On the retina, they form an irregular 2D tiling with an average distance of about $\simeq 50 \, \mu m$. The cover factor (CF) is defined as the number of synaptic trees which overlay a given geographic point on the retina. It is a quantity that is directly measurable by observation.  It depends on the distance to the visual streak and varies from 30 to 70  \cite{tauchi-masland:84}. The cover factor is proportional to  the ratio between the radius of the synaptic tree and the average distance between two closest cells, but the proportionality coefficient depends on the details of the statistical distribution of the cells. Nevertheless, a CF of $\simeq 30$ is compatible with the possibility for a given cell to be connected with almost one hundred other neighbouring cells (in agreement with the $\simeq 20$ simultaneously active synaptic releases experimentally observed in \cite{zheng-lee-etal:06}.

This rather large number of potential neighbouring cells greatly increases the number of operations to be performed at each time step of the numerical simulation. Combined with the huge difference between the fastest and slowest time scales in (\ref{eq:FullModel}), this finalises in making this type of numerical investigations especially time consuming. Therefore, for the sake of rigour and precision and although some 2D exploratory numerical simulations have been performed (see conclusion section), here we will restrict to 1D configuration. We are well aware that the choice of this geometry tends to move us away from reality, but the physical mechanisms bring into play, especially those associated with the coherence resonance phenomena, are still valid and then properly described.

We used a fourth order Runge Kutta algorithm with a $\simeq 0.1 ms$ time step. Several number of cells ($nc$) have been used, from 128 to 2048, although the vast majority of the simulation deals with $1024$ cells. We assume that each cell is symmetrically connected to the nearest  $2 \,\, n$ neighbours,  $n$ on each side. In our 1D simulations, $n$ change from $3$ to $95$, but most of the results deal with $n=12$. Finally periodic, but also vanishing and zero flux boundary conditions have been employed without relevant modification in the results.

\section{Synchronisation phase transition }\label{Sec:SynchroPhase}

\begin{figure}
\resizebox{0.5\textwidth}{!}{
\includegraphics[]{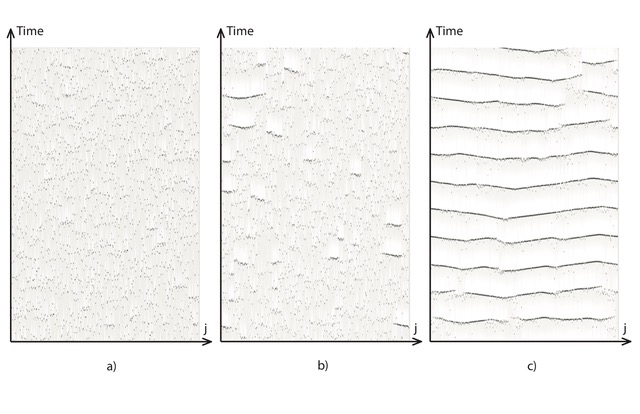}
}
\caption{Spatio-temporal evolution of the calcium concentration $C_{j}(t)$ (highest values are in dark). We choose black and white representation instead of colors for a better legibility of the figure. Cell indexes correspond to the horizontal axis, time is increasing along the upward vertical axis. The simulation involves 1024 cells, arranged in a circle, with 24 synaptically connected nearest neighbors for each cell. $g_{H}=10 \, nS$. The total record time is $880 \, s$, and the periodic sample is $1.1 \, s$. From left to right,  $g_{A}=6 \, pS$ (a), $8 \, pS$ (b) and $9 \,pS$ (c).}
\label{Fig:XT}
\end{figure}

Fig. \ref{Fig:XT} displays typical spatio-temporal evolutions of $C_{j}(t)$, the intracellular calcium concentration of cell $j$, for various values of $g_{A}$. The horizontal axis corresponds to $j$ while the vertical axis stands for the time $t$. Because of the huge difference between typical time scales, not all the time evolution is shown but only a periodic sampling ($1$ frame per $1.1 \,s$). Nevertheless, spontaneous propagating calcium waves are clearly visible, corresponding to bursts temporary synchronized  between neighbours. For small values of $g_{A}$ (fig. \ref{Fig:XT} a), these synchronisations events are short both in time and in the number of cells involved, but they clearly become longer with increasing value of the coupling strength $g_A$ (fig. \ref{Fig:XT} b and c). After the passage of an avalanche, one sees clearly the SAHP-driven refractory period lasting several seconds. This is particularly visible in fig. \ref{Fig:XT} b where three dynamical regimes can easily be distinguished: thin black lines (avalanches), followed by a pure white area (rest state) and finally a rather homogeneous "grey" zone associated with noise driven isolated spiking.  For high value of $g_{A}$ (fig. \ref{Fig:XT} c), the refractory period (pure white) imprints the network so that waves (grey) are not possible during a long period.

 \subsection{Firing Rate}
 \label{Sec:FiringRate}
We qualify a cell $j$ as "active at time $t$" if the calcium concentration $C_{j}(t)$ is higher than a given threshold concentration $C_{th}$. At rest, the concentration is close to $C_{0}$ but it can jump to $4$ or $5$ times $C_{0}$ in a burst. This provides a quantitative criterion to define an "efficient burst", namely a burst that lasts long enough so that sAHP rising and acetylcholine production (of the same time scale as calcium production) can take place. Arbitrary, we chose $C_{th}=2 C_{0}$. With this choice, the bursts are detected for sure but some isolated noise driven activities can also be detected. At a given time $t$, the firing rate $F\!R(t)$ is then defined as the ratio between the number of active cells and the total number of cells ($nc$).

\begin{figure}
\resizebox{0.50\textwidth}{!}
{
\includegraphics[]{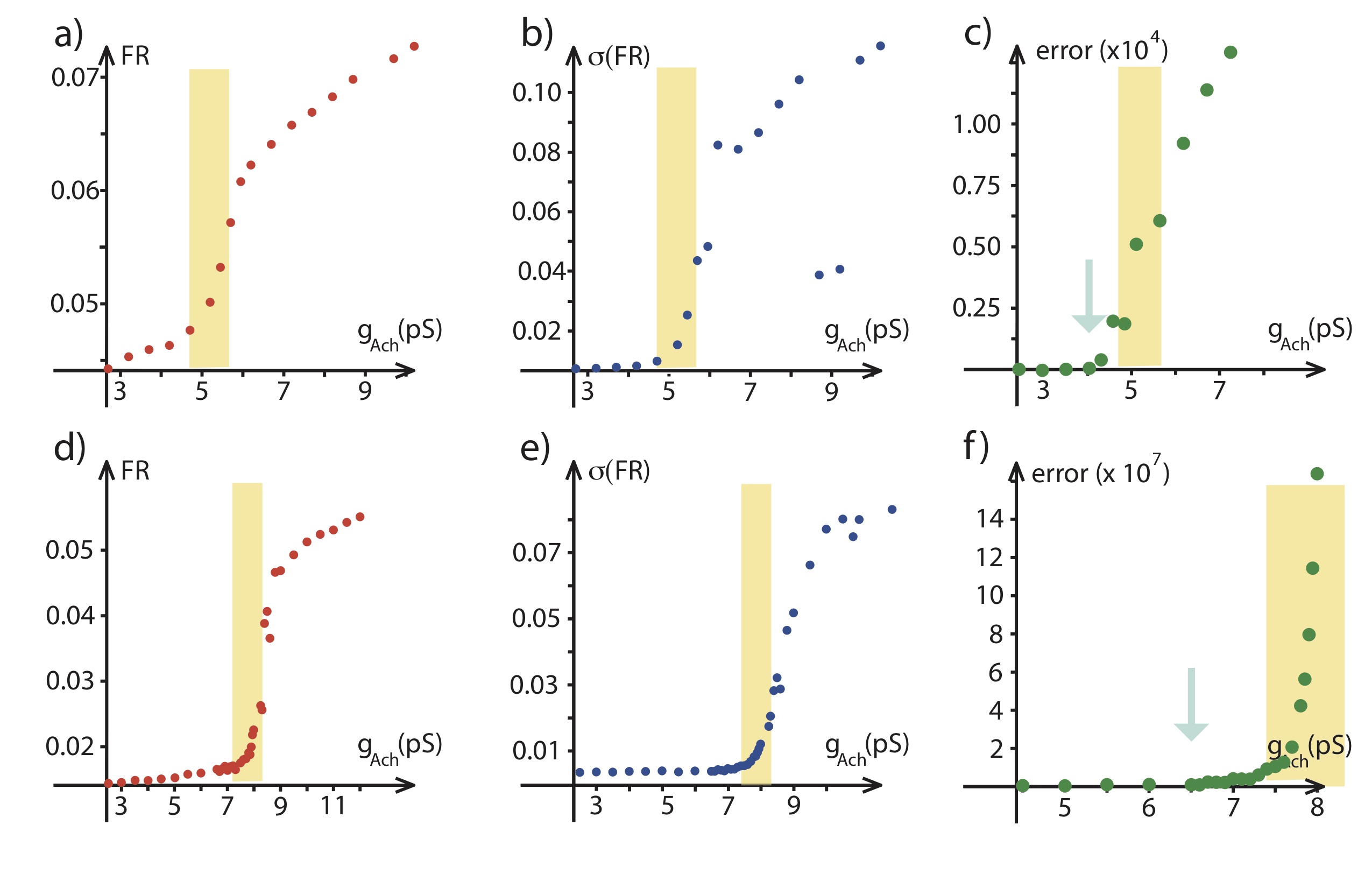}
}
\caption{Numerical simulation showing the evolution of $\av{F\!R}$ and $\sigma_{F\!R}$ with $g_{A}$. The top figures (a, b, c) are associated with  $ g_{H}=4 nS $ while the bottom ones (d, e, f) with $ g_{H}=10 nS $. The simulation involves 1024 cells, arranged on a circle, with 24 synaptically connected nearest neighbours for each cell. The first column (a, d) shows the average value of the firing rate versus $g_{A}$. The pale orange rectangular window points to the transition region. Its position is determined with naked eye and its width stands for the measurement uncertainty. The second column (b, e) shows the corresponding standard deviation of the firing rate versus $g_{A}$. The transition region determined in column (a, d) has been identically reported in (b,e) and still matches with the area where a strong change takes place. Finally in the last column (c, f), we have reported  $\epsilon$ (definition \eqref{eq:epsilon} in the text ) versus $g_{A}$. Again the transition area identified in (a, d) or (b, e) has been identically reported and does correspond to a strong $\epsilon$ variation. However the figures (especially f) clearly show that $\epsilon$ actually starts to increases a little bit before the strong transition region spotted by the pale orange window (vertical blue arrow). Finally, note in b two points close to $g_{Ac}\simeq 9 \, pS $, that stand out and depart clearly from the alignment with the other experimental measurements. An explanation is given later in the text.
\label{Fig:FRSigmaError}
}
\end{figure}

In fig.  \ref{Fig:FRSigmaError} we display the mean of $F\!R(t)$, $\av{F\!R}$ and its mean square deviation $\sigma_{F\!R}$, as a function of 
$g_{A}$. The most obvious observation is then the existence of a steep increase of both the average and standard deviation in the neighborhood of $g_{A}\simeq 5.2 \, pS$ (for $g_{H}=4 \, nS$) and $g_{A}\simeq 7.5 \, pS$ (for $g_{H}=10 \,nS$).
Calling $g_{A}^{n}$ these naked-eye estimated threshold values, we indicate their location in fig.\ref{Fig:FRSigmaError} by an orange window, centered on the threshold value and whose width stands for an estimate of the measurement error. They clearly distinguish between two phases. On the left, the activity is essentially noise driven with a low probability that calcium concentration exceeds the threshold. In this phase $F\!R$ is not exactly vanishing for small values of $g_{A}$.  This residual value depends on $C_{th}$ and can be reduced by increasing it. The activity in this region is almost homogeneous with respect to time and space, and therefore leads to small fluctuations of the $F\!R$ in the neighbourhood of its average value. On the opposite, the region on the right of the orange window corresponds to the presence of a intermittent collective behaviour nicely characterized by $\sigma_{F\!R}$ (fig. \ref{Fig:FRSigmaError}, column 2). Synchronised dynamics leads to bursts of activity, strong temporal variations of $F\!R$ and then to large value of $\sigma_{F\!R}$. 

In order to study in more detail the transition area, we introduce the $\epsilon(g_{A})$ function defined as follows. Note that $\sigma_{FR}$ is a function of $g_A$. For a fixed value of $g_{A}$ we estimate the shape of the curve $S_{g_{A}}=\left\{ (g,\sigma_{F\!R}(g)) , \,  g \leqslant g_{A}\right\}$  by a second order polynomial $P_{g_{A}}$. $S_{g_{A}}$ is the curve one can fit knowing the values of $\sigma_{F\!R}(g_{A})$ for Ach conductances smaller than $g_{A}$. Let $n(g_{A})$ be the number of points in $S_{g_{A}}$. We introduce:
\begin{equation}\label{eq:epsilon}
\epsilon(g_{A})={{1}\over{n(g_{A})}} \sum_{x \in S_{g_{A}}} \left(P_{g_{A}}(x)-\sigma_{F\!R}(x)\right)^2,
\end{equation}
which measures  how much (in the $L^2$ sense) the standard deviation $\sigma_{F\!R}(g_{A})$ deviates from the value that can be guessed from smaller values of $g_{A}$. As long as $\sigma_{F\!R}$ evolves regularly with $g_{A}$, then its behavior for values a little larger is predictable and the difference, measured by $\epsilon$, between the real value and that predicted, is small. On the contrary, larger values of $\epsilon$ indicates the presence of a brutal change. In fig. \ref{Fig:FRSigmaError} c and \ref{Fig:FRSigmaError} f, we expect and we do observe a strong variation of $\epsilon(g_{A})$ in the neighbourhood of $g_{A}^{n}$. We checked this observation persists even when the interpolation polynomial degree is increased above $2$.

\subsection{Distribution of the avalanche size}
 \label{Sec:PDF}
With regard to numerical simulations, a precise definition of what we call an "avalanche" can be formulated, with as few arbitrary parameter choices as possible.  There are $nc$ cells with periodic boundary conditions and the numerical simulation implies the existence of a small enough time increment $dt$. We introduce the state variable $a(j,p)$ equals to $1$ if the calcium concentration at cell $j$ at time $p \, dt$ is higher than $C_{th}$ and equal to $0$ otherwise.
Two "on" states  $a(j,p)=1$ and  $a(j',p')=1$ are said to be connected if and only if there exist a continuous path of excited cells connecting them. The avalanche associated with the "on" cell $a(j,p)=1$ is then the set of all the "on" cells which are connected with it. Therefore, in our definition of avalanche, there is a notion of cells connexity and propagation, in contrast to other works where this notion is not taken into account. Our definition indeed takes carefully into account the wave mechanism underlying these avalanches which involves a topological structure of neighbouring cells.

\begin{figure}
\resizebox{0.50\textwidth}{!}{
\includegraphics[]{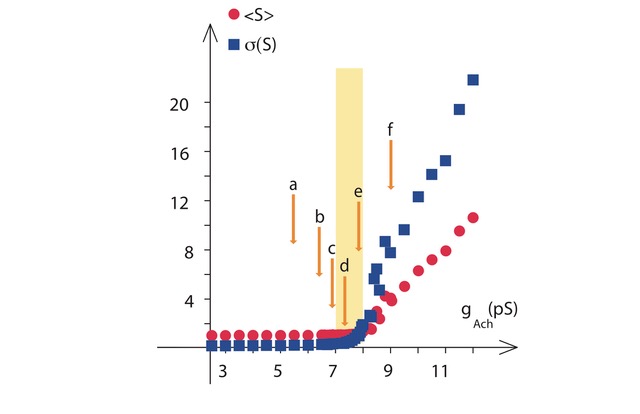}
}
\caption{Numerical simulation with the same parameters as in fig. \ref{Fig:FRSigmaError} (bottom, $g_{H}=10 \, nS$), showing the evolution of the mean (red circle) and standard deviation (blue squares) of the avalanche size distribution versus $g_{A}$. The arrows point to specific values of $g_{A}$  used in  fig. \ref{Fig:PDF}.}
\label{Fig:PowerExpDistance}
\end{figure}

We first measure how the average and  standard deviation of the avalanche size distribution evolve with increasing synaptic coupling strength $g_{A}$. As for the firing rate in fig. \ref{Fig:FRSigmaError} bottom, fig. \ref{Fig:PowerExpDistance} clearly reports on a brutal behaviour change in the avalanche size distribution around $g_{A} =g_{A}^{n} \simeq 7.5 \, pS$ for $g_{H}=10 \, nS $. It is worth observing that, above the pale orange rectangular window which stands for $g_{A}^{n}$ and its measurement uncertainty, the standard deviation of the avalanche size distribution is not only greater, but also  increases much faster than the average size.  Although this remark alone does not allow to identify the shape of the avalanche size distribution curve, it nevertheless goes in the direction of a scale free process, i.e. without characteristic size.

\begin{figure}
\resizebox{0.5\textwidth}{!}{
\includegraphics[]{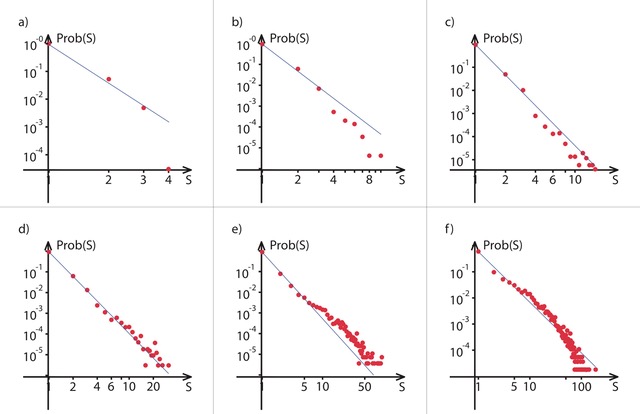}
}
\caption{Probability density of the avalanche sizes in log-log scale. Note that the horizontal scale changes significantly from one plot to another. The red disk are numerical measurements while the blue line corresponds to the best power law fit. All the parameters, except $g_{A}$ are the same as  in fig. \ref{Fig:XT}. From top to bottom and from left to right: a: $g_{A}=5.5 \, pS$,  b: $g_{A}=6.5 \, pS$,  c: $g_{A}=7.0 \, pS$,  d: $g_{A}=7.5 \, pS$,  e: $g_{A}=8.0 \, pS$,  f: $g_{A}=9.0 \, pS$. These values are labelled on  fig. \ref{Fig:PowerExpDistance} and on fig. \ref{Fig:PowerExpDistance2} with vertical orange.}
\label{Fig:PDF}
\end{figure}

Avalanche size distributions are shown in Fig. \ref{Fig:PDF}. The various values of the synaptic coupling strength $g_{A}$ labelled on fig. \ref{Fig:PowerExpDistance} with vertical orange arrows have been chosen in order to frame the zone around $g_{A}^{n}$.   We use log-log scales for a straightforward comparison with possible power law curve. The evolution of both the average firing rate (fig. \ref{Fig:FRSigmaError} d,e) and  the standard deviation of the avalanche size (fig. \ref{Fig:PowerExpDistance}), already provides a coherent possible value of the $g_{A}$ synchronisation threshold close to $g_{A}^{n}=7.5 \, pS$ for $g_{H}=10 \, nS$. We now show that there is a value of $g_{A}$, depending on $g_{H}$ and $\eta$, where the waves distribution is the closest to a power law, where the distance is measured by the Batthacharyya's distance.

For two probability distributions $p$ and $q$ defined over the same domain $X$, the Batthacharyya's distance between $p$ and $q$, noted $d_{B} \left( p,q \right)$, is defined as
\begin{equation}
d_{B} \left( p,q \right) =-ln \left(  \sum_{x \in X}  \sqrt{p(x)q(x)}\right)
\end{equation}
The terminology "distance" is improper because $d_{B}$ does not satisfy the triangular inequality. However, the  Hellinger's distance $d_{H}=\sqrt{1-e^{-d_{B}}}$, which is monotonously related to Batthacharyya's one, does. Finally, note also that i) the Batthacharyya's distance depends mainly on the values of $x$ that are most likely ii) the distance is defined even for small statistical set $X$ (as in fig.\ref{Fig:PDF}a).

\begin{figure}
\resizebox{0.50\textwidth}{!}{
\includegraphics[]{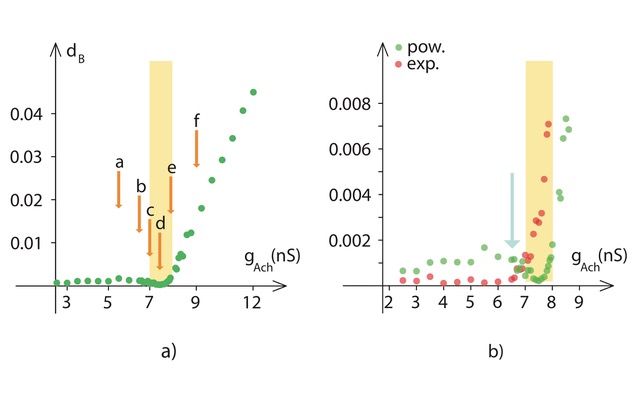}
}
\caption{ Numerical simulation of eq. (\ref{eq:FullModel}) with the same parameters as in fig.\ref{Fig:XT}, showing a: Bhattacharya's distance between the numerical probability density function of the avalanche size and its best power law fit, b: Comparaison between the distance to the best power law fit (green dots, vertical enlargement of a) and the distance to the best exponential fit (red dots). We have identically reproduced the orange window and vertical blue arrow previously introduced in fig.\ref{Fig:FRSigmaError}.}
\label{Fig:PowerExpDistance2}
\end{figure}

\begin{figure}
\resizebox{0.45\textwidth}{!}{
\includegraphics[]{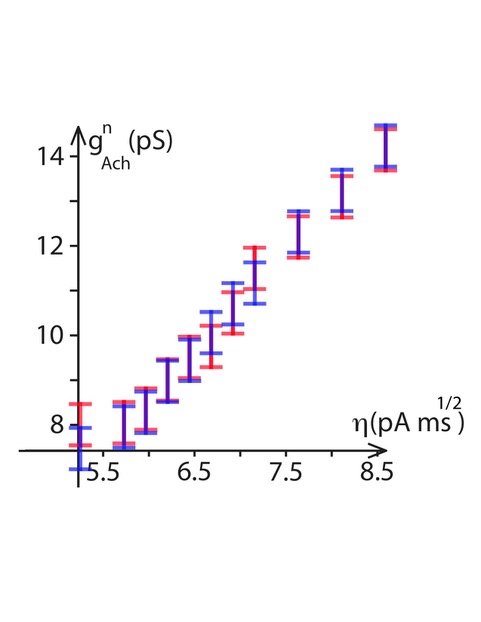}
}
\caption{Coincidence between two independent measurements of $g_{A}^{n}$ versus the noise amplitude $\eta$. The parameters are the same as in fig. (\ref{Fig:XT}).
In the text, two measurements of $g_{A}^{n}$ have been reported. In section \ref{Sec:FiringRate}, $g_{A}^{n}$ is first introduced as the value of the coupling strength value associated with a sudden increase of the average and standard deviation of $F\!R$ (vertical red bars). Then,  in section \ref{Sec:PDF}, $g_{A}^{n}$ was identified as the coupling strength value for which the probability density function of the avalanche size looks most like a power law (vertical blue bars).  The vertical bar length stands for the measurement error. As the figure shows, the two measurements accurately coincide.} 
\label{fig2SeuilVsNoise}
\end{figure}

Fig. \ref{Fig:PowerExpDistance2}a shows the Batthacharyya's distance between the numerically measurement of the avalanche size distribution ($P_{exp}(S)$, $ S \ge 1$) and the best power law fit $P_{b}(S)={{ \zeta(b) }\over{S^b}}$, where $\zeta(b)$ is a normalisation constant depending on the exponent $b$. For each value of $g_{A}$, we first compute $d_{B}(P_{exp},P_{b})$ and then optimize the outcome with respect to $b$. The curve $d_{B}(P_{exp},P_{b})$ as a function of $g_{A}$ displays a minimum for $g_{A}=g_{A}^{n} \simeq 7.5 \, pS$ ($g_{H}=10 \, nS$). This is in very good agreement with the previous and independent measurements of the coupling threshold $g_{A}^{n}$, for which a divergence of the correlation length is expected.  In fig. \ref{fig2SeuilVsNoise} we check this coincidence for several values of the noise amplitude $\eta$. Finally fig. \ref{Fig:PowerExpDistance2}b displays, not only the evolution of distance to the best power law  - vertical enlargement of fig. \ref{Fig:PowerExpDistance2}a -, but also the distance to the best exponential fit. 

Both the measurements of the average firing rate (fig.\ref{Fig:FRSigmaError}), of the average and standard deviation of the avalanche size (fig.\ref{Fig:PowerExpDistance}), and of the Bhattacharya's distance to the best power fit (fig.\ref{Fig:PowerExpDistance2}), independently contribute to prove the existence of a transition synchronisation threshold close to $g_{A}^{n}$. Now, besides $g_{A}^{n}$, a more subtle observation of fig.\ref{Fig:FRSigmaError}c,f and fig.\ref{Fig:PowerExpDistance2}b, reveals  the existence of an other noteworthy $g_{A}$ coupling value, smaller than $g_{A}^{n}$ and materialized in the figures by a vertical pale blue arrow. We call this value $g_{A}^{p}$. This same value is associated with the coupling value for which: i) in fig.\ref{Fig:FRSigmaError}c,f $\epsilon(g_{A})$ starts to deviate from zero, ii) in fig. \ref{Fig:PowerExpDistance2}b, the Bhattacharya's distance to an exponential law starts to increase.

\section{travelling waves and their nucleation}\label{Sec:TravellingWavesNucleation}
Up to now, for $g_{H} =10 \,nS$, we have reported on the existence of two noteworthy synaptic coupling values: $g_{A}^{n}\simeq 7.5\, pS$ (pale orange rectangular window) associated with the synchronisation threshold, and $g_{A}^{p}\simeq 6.5 \,pS$ (vertical blue arrow) which looks like a precursor. In what follows, we are going to propose a theoretical interpretation of these noticeable values. 

\begin{figure}
\resizebox{0.5\textwidth}{!}{
\includegraphics[]{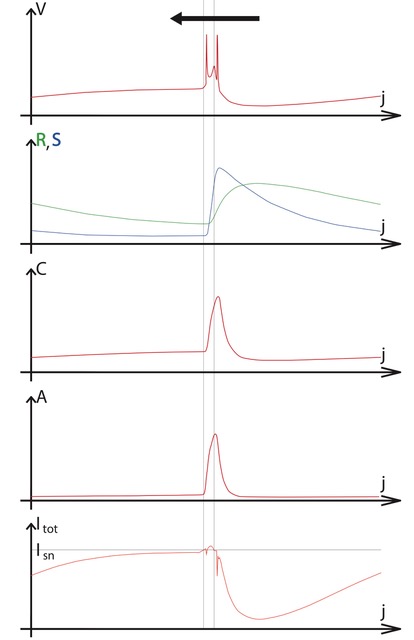}
}
\caption{ Travelling wave solution. The parameters are the same as in fig.\ref{Fig:XT} c, except that here the noise amplitude is vanishing. The horizontal axis corresponds to the cell indexes $j$, from $1$ to $512$, with periodic boundary conditions. The arrow stands for the propagation direction.  From top to bottom, $V$, $R$ (green) and $S$ blue), $A$ and $C$ and finally $I_{tot}$ (see eq. \ref{eq:FastEquations}). For $I_{tot}$ plot, the horizontal line corresponds to $I_{SN}$. The vertical lines across the five plots are associated with the first and last crossings $I_{tot} =  I_{SN}$. }
\label{figSoliton}
\end{figure}

\subsection{Travelling waves}\label{Sec:TravellingWaves}
In the absence of noise ($\eta \simeq 0$) and with periodic boundary conditions, eq. (\ref{eq:FullModel}) possesses travelling wave solution. They are classical excitable wave \cite{bressloff-coombes:98}, periodic in space and in time, propagating, in 1 D,  either on the left or on the right and disappearing by collision. Here the refractory tail is associated with the SAHP mechanism.  Fig. \ref{figSoliton}
displays a period of the typical solution profile. The stability of the travelling waves has been investigated numerically. As for usual excitable waves, the travelling solution may become unstable when its wavelength is not large enough \cite{meron:92}. Indeed in such a case, each travelling excitable pulse catches the refractory tail of the previous pulse and stops. For large enough wavelength, the stability depends on the acetylcholine coupling between the cells: the travelling solution is stable for large $g_{A}$ but stops to propagate and disappears as soon as $g_{A}$ is smaller than a well defined threshold. The mechanism can be deeper understood through a rough analytical computation displayed in \ref{AppendixThreshold}.

\begin{figure}
\resizebox{0.5\textwidth}{!}{
\includegraphics[]{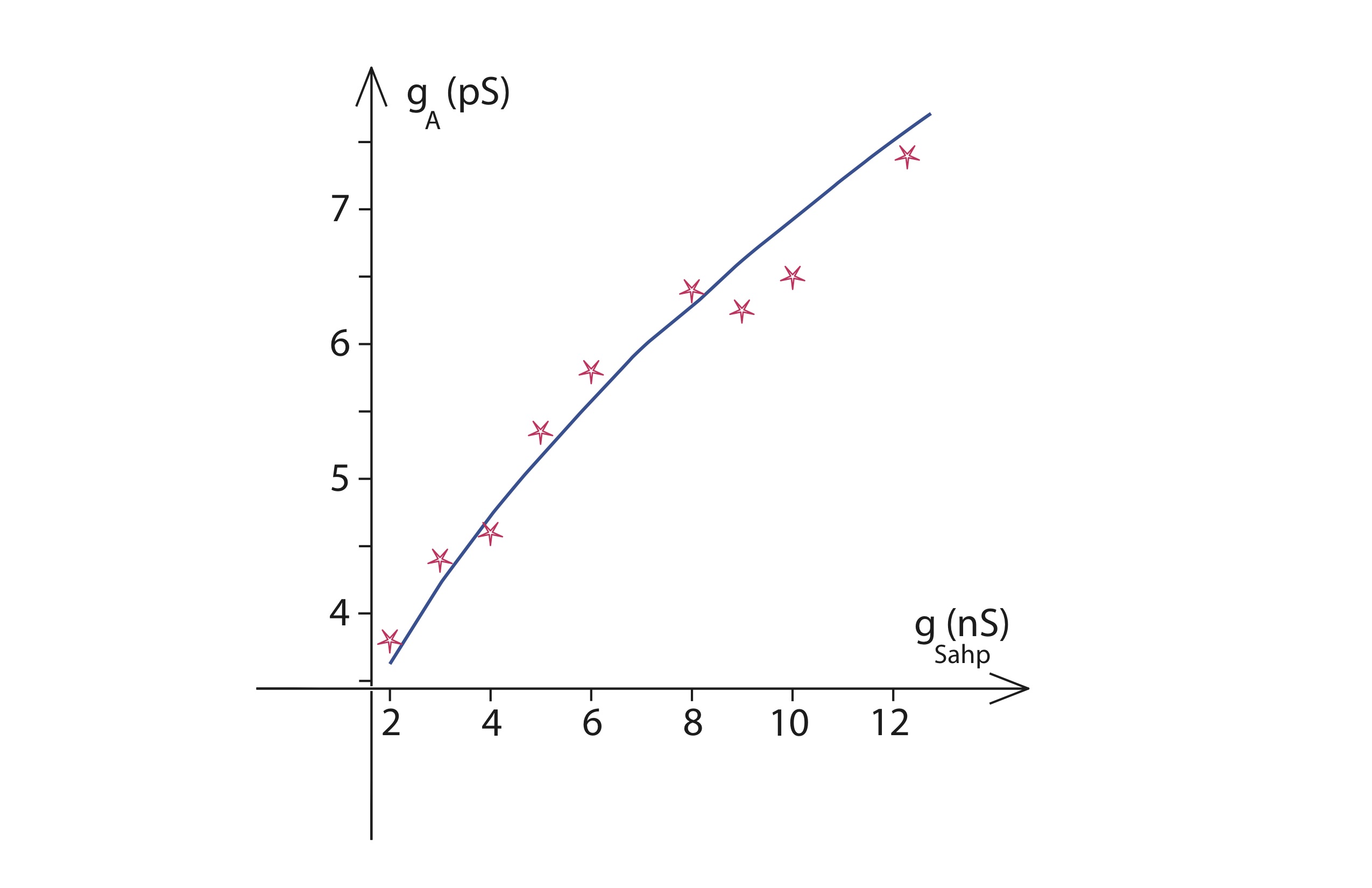}
}
\caption{The parameters are the same as in fig.\ref{Fig:XT}c except that here the noise amplitude is vanishing and $g_{H}$ is varying. The red asterisks correspond to numerical determination of $g_{A}^{p}$ from the observation of $\epsilon(g_{A})$ curves (like those in fig.\ref{Fig:FRSigmaError}c). The continuous blue curve is determined by accurate numerical investigations of the travelling wave stability in absence of noise. Above this blue line, the travelling waves are stable.}
\label{figSolitonThreshold01}
\end{figure}

We have accurately measured the threshold associated with the travelling wave stability in absence of noise for several values of $g_{H}$ (continuous blue line in fig.\ref{figSolitonThreshold01}), and do observe that it is definitely different from $g_{A}^{n}$. By cons, it accurately coincides with $g_{A}^{p}$ (red dot asterisks in fig.\ref{figSolitonThreshold01}), the smallest noteworthy $g_{A}$ coupling strength related to the beginning of occurrence of non vanishing $\epsilon(g_{A})$ values and the decline of the best exponential fit. We interpret this result in the following way: At each time step, the set of all SACs is described by a state vector $W=\left[V_{j},N_{j},R_{j},S_{j},C_{j},A_{j}, j \in [1,n_{c}] \right]$ which lives in  a compact subset of $\mathbb{R}^{6 n_{c}}$ space. With time and in absence of noise, $W$ describes a deterministic trajectory whose starting point is given by the initial conditions $W(t=0)$. A travelling wave solution is then associated with a periodic orbit in this space, and the set of all the initial conditions which asymptotically converge to this periodic orbit is called the attraction basin of the travelling wave. Hence by definition, attraction basin borders are separatrix of the dynamics, and the basin of attraction of an unstable solution has zero volume. The presence of the noise leads to a broadening of deterministic trajectories. Especially the borders of the attraction basin are no longer insurmountable boundaries: they can be crossed thanks to noise. Hence, an initial condition, which in the absence of noise would have led to a travelling wave, can, in presence of noise, give birth to a trajectory that will persist for a time in the travelling wave basin of attraction, and finally end up escaping, resulting in a spatio-temporal structure that looks like a travelling wave embryo, i.e. something like an unsuccessful or an incomplete wave start. Conversely, an initial condition outside the attraction basin, which in the absence of noise would not have led to a travelling wave, can, in presence of noise cross the attraction basin border and looks like a travelling wave for a moment. We make the assumption that these finite duration waves are in fact what is called avalanche. In this scenario, the existence of avalanches requires at least the existence of attraction basin, i.e. stability of the traveling wave and $g_{A}>g_{A}^{p}$.

\subsection{Nucleation of travelling waves} \label{Sec:Nucleation}
Hence above $g_{A}^{p}$, the dynamics starts to be sensitive to the presence of a travelling wave attraction basin. But this is not the end of the story, because two distinct and somewhere competing mechanisms, both driven by the noise fluctuations, are now bring into play. 
On the one hand, noise fluctuations can help the system to jump from its rest state into the traveling wave attraction basin and are therefore directly involved in the frequency with which avalanches are generated. On the other hand, noise fluctuations are also straightforwardly involved in the time spent by the trajectory in the travelling waves attraction basin as well as the minimal distance at which the trajectory approaches the exact travelling orbit. It is no longer a question of wave generation frequency, but rather a matter of quality of the generated waves.

\begin{figure}
\resizebox{0.5\textwidth}{!}{
\includegraphics[]{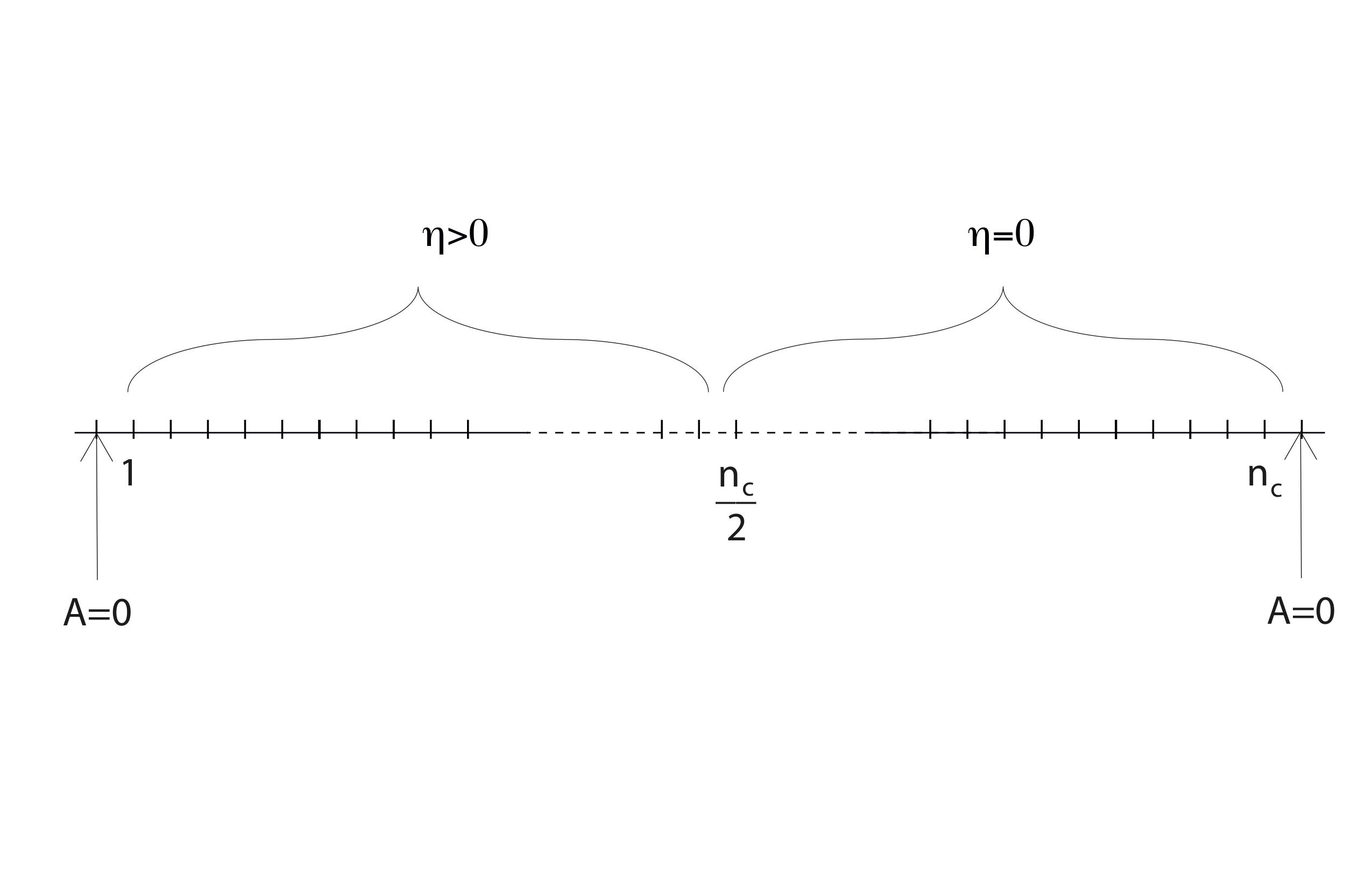}
}
\caption{Schematic representation of the additional numerical simulation performed in section \ref{Sec:Nucleation}. Vanishing boundary conditions are imposed for both left and right ends. The cells on the left, from $0$ to ${{n_{c}}\over{2}}-1$ are subjected to noise fluctuations ($\eta>0$), unlike the rest of the box, from ${{n_{c}}\over{2}}$ to $n_{c}$ for which $\eta=0$.}
\label{figSolitonThreshold02}
\end{figure}
In order to investigate this last point, a new experiment in which the detection of a travelling wave is done for sure, has been designed. We use the same parameter regime as in  fig. \ref{Fig:XT} but with specific boundary conditions and inhomogeneous repartition of the noise amplitude (see diagram fig.\ref{figSolitonThreshold02}). We impose vanishing boundary conditions for both left and right ends (i.e. $ A=0$). Also the cells on the left half side (from $0$ to $nc/2-1$) are submitted to noise while those on the right half side (from $nc/2$ to $nc$) are not ($\eta=0$). We use values of the synaptic coupling $g_{A}$ slightly higher than $g_{A}^{p}$ such that, in absence of noise, periodic travelling wave, once created, can propagate. Therefore, in the right side of the numerical box, cells tend to stay in their rest state (i.e. waves do not spontaneously occur) except when waves created in the left side propagate through the right side and disappear at the right end. Hence this right side plays the role of a travelling wave revealer, i.e. a place where the traveling wave can be undoubtedly identified.

\begin{figure}
\resizebox{0.5\textwidth}{!}{
\includegraphics[]{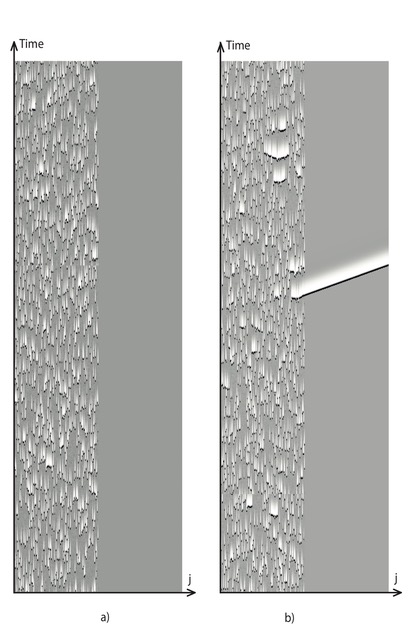}
}
\caption{Typical spatio-temporal evolution of the calcium concentration $C_{j}(t)$ (highest values are in dark). Cell indexes j are on the horizontal axis, time is increasing along the upward vertical axis. The parameters are the same as in fig.(\ref{Fig:XT}) but the geometry corresponds to fig.(\ref{figSolitonThreshold02}). a) $g_{A}=7.3 \, pS <g_{A}^{n}$. b) $g_{A}=8.0 \, pS >g_{A}^{n}$. For a) the lack of travelling wave in the right side persists even for very long time}
\label{figNUC}
\end{figure}

\begin{figure}
\resizebox{0.5\textwidth}{!}{
\includegraphics[]{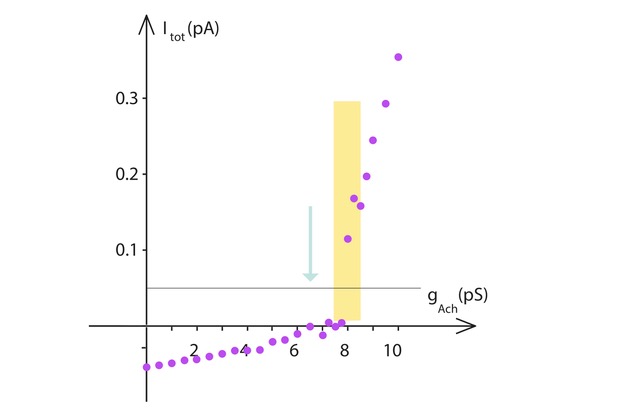}
}
\caption{ Maximum value of $I_{tot}$ measured at cell $nc/2$ (see text) versus $g_{A}$. The horizontal black line stands for $I_{SN}$. The parameters are the same as in fig.(\ref{Fig:FRSigmaError}), and the pale orange window as well as the vertical blue arrow have been identically reported from that figure.}
\label{Fig:SecondSeuil2}
\end{figure}

We then numerically observe that in the left side, the dynamics looks like those in fig. \ref{Fig:XT}, i.e. with the spontaneous occurrence of spatio-temporal areas of synchronisation (fig.\ref{figNUC}). Therefore, for long enough recording time and for $g_{A}> g_{A}^{p}$, it always happens that a synchronisation area occurs close to the $nc/2$ border, with a velocity toward the right side. But, and this is the crucial point, that this burst of synchronisation is able to give birth to a travelling wave propagating in the right side (i.e to {\it nucleate} a traveling wave), depends on whether $g_{A}$ is smaller (fig.\ref{figNUC}a) or higher (fig.\ref{figNUC}b) than $g_{A}^{n}$. This qualitative observation is supported by the measure of the external current  seen by the cell number $nc/2$ (i.e. $I_{tot}(nc/2)$), especially by the maximum value reached during a sufficiently long observation period (fig.\ref{Fig:SecondSeuil2}). In this figure, we do observe the existence of a nucleation threshold close to $7.5 \, pS$,  i.e. exactly the same value as the threshold observed for the firing rate evolution $g_{A}^{n}$. 
We checked this coincidence for several values of the noise amplitude (not shown).

We note that, for large enough coupling strength and even in presence of noise (provided $\eta$ is small enough), travelling waves can propagate without being destroyed. The occurrence of these steady travelling waves for large values of $g_{A}$ depends on the initial conditions, so they may be present or not, but once created, they persist. In fig.\ref{Fig:FRSigmaError}b, the two unexpected 
points close to $g_{A} \simeq 9 \, pS$ and $\sigma_{F\!R} \simeq 0.04$ correspond to the accidental presence of such steady travelling waves.  Hence, besides $g_{A}^{p}$ and $g_{A}^{n}$, there also exists a third $g_{A}$ threshold, larger than the previous ones, even if, for the present study, we are not concerned with it.

\section{Role of the noise}\label{Sec:RoleNoise}

\begin{figure}
\resizebox{0.50\textwidth}{!}{
\includegraphics[]{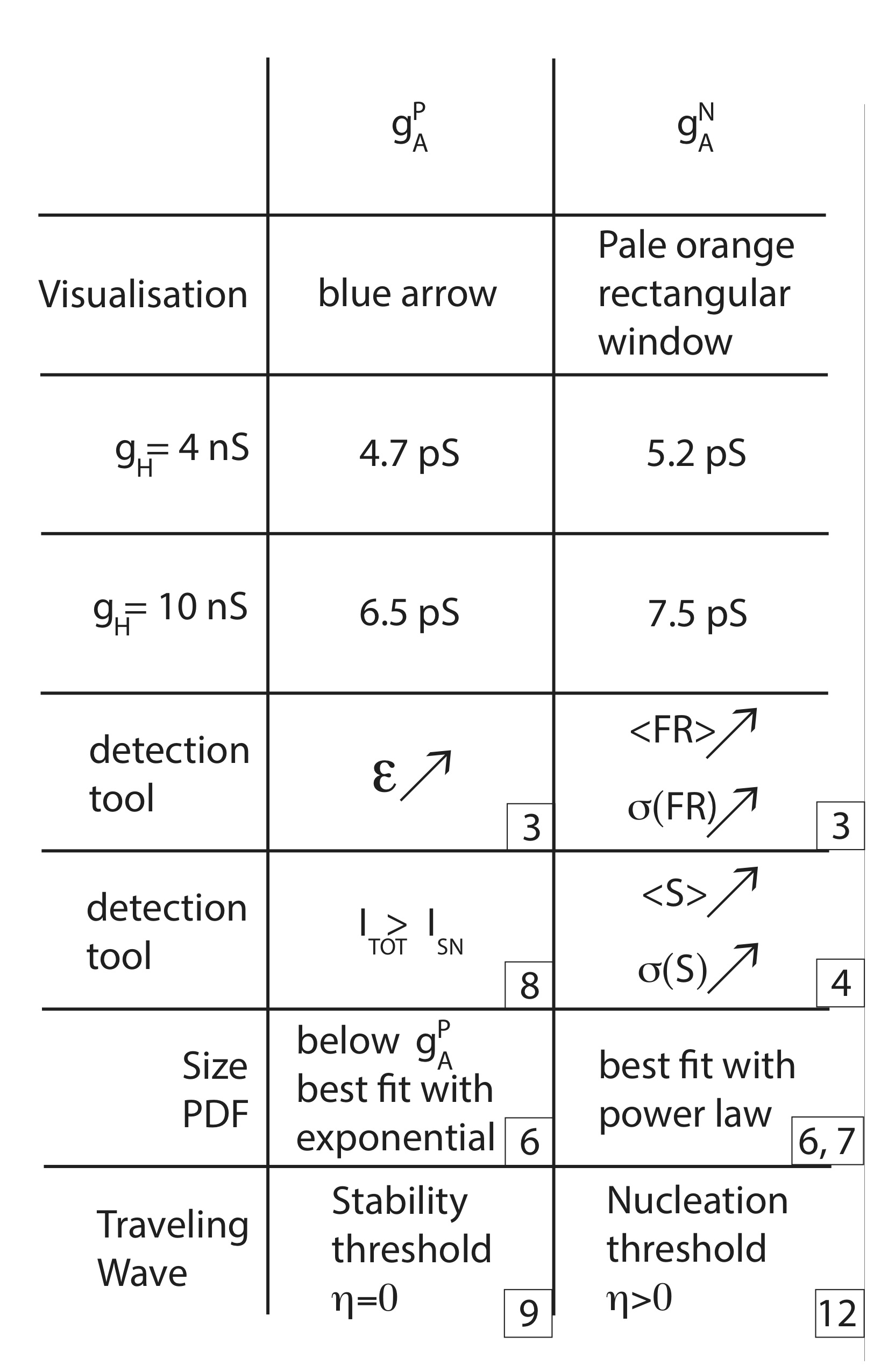}
}
\caption{Summary table of $g_{A}^{p}$ and $g_{A}^{n}$ characteristics and properties. Oblique arrows indicate the presence of a sudden variation. The numbers in the boxes at the bottom right correspond to the number of figures.}
\label{Fig:recapitulatif}
\end{figure}

Up to now, we have identified two noticeable coupling strength values: $g_{A}^{p}$ associated with the travelling wave stability threshold in the absence of noise, and $g_{A}^{n}$ related to the spontaneous nucleation of travelling waves in the presence of noise. Tab.\ref{Fig:recapitulatif} presents a summary of their respective properties.

\begin{figure}
\resizebox{0.5\textwidth}{!}{
\includegraphics[]{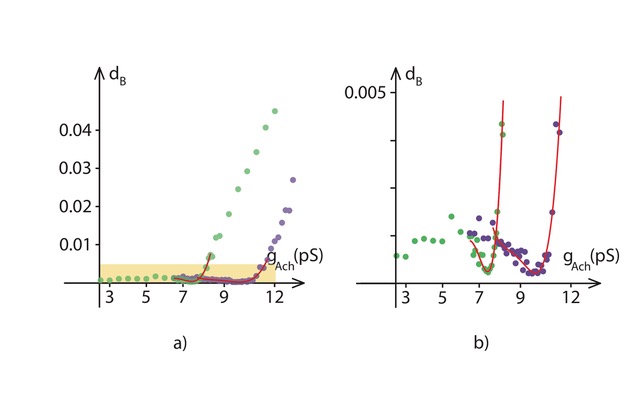}
}
\caption{Bhattacharya's distance between the numerically measurement  of the probability density function of the avalanche size and its best power law fit, for two distinct noise amplitude: $5.25 \, pA \, ms^{{1}\over{2}}$ (green) and $6.68 \, pA \, ms^{{1}\over{2}}$ (purple). The continuous red lines correspond to 4th order polynomial fit in the neighborhood of the local minimum.  b: is an enlargement of the orange rectangle in a} 
\label{Fig:BatFor2Noise}
\end{figure}

We now investigate the effect of noise onto the synchronisation transition. We proceed in the following way: the regime of parameters is the same as in fig. \ref{Fig:XT} except that the noise amplitude $\eta$ varies from $5.25 \, pA \, ms^{{1}\over{2}}$ to $8.59 \, pA \, ms^{{1}\over{2}}$. For each value of the noise amplitude, we scan several values of the coupling strength $g_{A}$, from far below $g_{A}^{p}$ to far above $g_{A}^{n}$. Then  for each pair $(\eta,g_{A})$, we  numerically measure the avalanche size probability density function $P_{exp}(S)$, compute $b^{*}$ which corresponds to the best  power law approximation $P_{b^{*}}(S)={{\zeta(b^{*})}\over{S^{b^{*}}}}$ and finally compute the Batthacharyya's distance between $P_{exp}(S)$ and $P_{b^{*}}(S)$.

Fig. \ref{Fig:BatFor2Noise} reports on this Bhattacharya's distance ($d_{B}$) for two distinct noise amplitudes. Let us call
$\eta_{-}$ the smallest one (green dots) and $\eta_{+}$ the highest one (purple dots).
Recalling that the nucleation threshold $g_{A}^{n}$ is associated with the minimum of $d_{B}$, we first observe that  $g_{A}^{n}(\eta_{-}) <  g_{A}^{n}(\eta_{+})$, in good agreement with the intuitive idea that the greater the noise, the harder it is to nucleate a long duration travelling wave. Fig. \ref{fig2SeuilVsNoise} confirms this trend for the range of noise amplitude $\bra{\eta_{-},\eta_{+}}$.

The second observation deals with the $d_{B}$ curvature $\rho$ at the minimum $g_{A}^{n}$. It is quantitatively determined from the computation of  the best polynomial fourth order fit (the continuous red lines in fig. \ref{Fig:BatFor2Noise})
\begin{equation}
d_{B}(g_{A}) \simeq d_{B}(g_{A}^{n})+{{1}\over{2}} \underbrace{{{\partial^2 d_{B}}\over{\partial g_{A}^2}} {\Big \vert}_{g_{A}^{n}}}_{\rho}\left(g_{A}-g_{A}^{n}\right)^2+...
\end{equation}
as its second order derivative. Dimensionally, the curvature $\simeq 1 / {\delta g_{A}}^2$ where $\delta g_{A}$ is a characteristic distance to the extremum $g_{A}^{n}$.  Here, already with naked eye, we observe that $\rho(\eta_{-})>\rho(\eta_{+})$, which means that the $g_{A}$ range of parameters for which the avalanche size pdf "looks like" a power law is larger for $\eta_{+}$ than for $\eta_{-}$.

\begin{figure}
\resizebox{0.5\textwidth}{!}{
\includegraphics[]{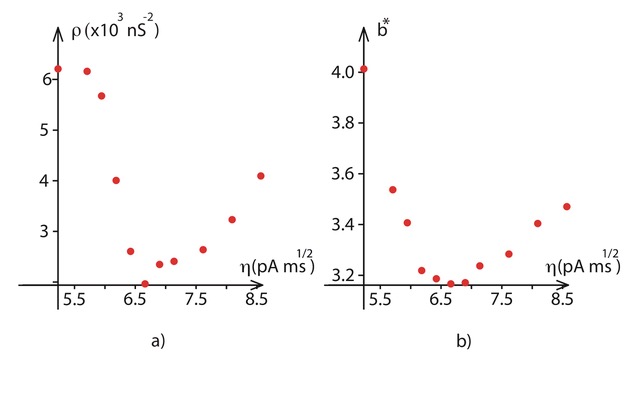}
}
\caption{a: Evolution of the curvature $\rho$ of the $d_{B}(g_{A})$ curve at the minimum $g_{A}^{n}$ versus the noise amplitude ($\eta$). b: Evolution of the exponent $b^{*}$ versus $\eta$.} 
\label{Fig:CourbureExposant}
\end{figure}

In Fig \ref{Fig:CourbureExposant}a we investigate the evolution of the curvature with respect to the noise amplitude. The evolution is not steadily varying but displays a minimum for a noise amplitude close to ${\widetilde \eta} \simeq 6.6 \, pA ms^{1/2}$. From our knowledge, this behaviour has never been reported before. What comes closest in the literature deals with the coherence resonance mechanism, i.e. with the existence of a noise amplitude for which the self correlation of an excitable system displays a maximum \cite{pikovsky-kurths:97}.  Later in \cite {kim-lee-etal:15} the relationship between coherence resonance and bursting neural networks has been investigated, with the experimental evidence that there is a noise strength for which the regularity of the interburst sequence becomes maximal. 
But here, we are dealing with a different mechanism: the broadening of the synchronization transition with noise. It means that the range of coupling strength around $g_{A}^{n}$ for which the avalanche size pdf looks like a power law, i.e. the neural system is critical, is maximal for ${\widetilde \eta}$. 

Fig.\ref{Fig:CourbureExposant}b reports on how the exponent $b^{*}$ of the best power law fit depends on the noise amplitude. We observe that $b^{*}$ displays a minimum close to $\simeq 3.17$ for the same noise amplitude ${\widetilde \eta}$ for which the curvature displays a minimum in fig.\ref{Fig:CourbureExposant}a. As for a power law $P_{b}(S)={{\zeta(b)}\over{S^b}}$, the smaller $b$ the larger the standard deviation, it means that not only ${\widetilde \eta}$ corresponds to the largest critical range of parameter $g_{A}$, but also to the largest distribution of avalanche size and therefore to the maximal sensitivity to stimuli. 
These observations form the main result of our study.

\section{Discussion and conclusion}
\label{Sec:Discussion}

\subsection{Self organized criticality}\label{Sec:SOC}
So far, in the studies of the mechanisms that allow a
neuronal system to self-organize in a critical state, noise was not expected to play a relevant role. The literature reported on two main mechanisms. Either the presence of a feedback loop which dynamically maintains the parameters controlling the synchronization in the neighborhood of its transition threshold \cite{sornette-johansen-etal:97, levina-herrmann-etal:07, di-santo-villegas-etal:18, kossio-goedeke-etal:18}, or the presence of a particular synaptic network with hierarchical modularity, small worldness and economical wiring \cite{rubinov-sporns-etal:11}. In both cases, noise was not playing any fundamental role.
Our results do not contradict these previous studies but the mechanism we propose is of a different nature: noise enlarges the critical range of coupling parameters near the saddle-node bifurcation leading cells to burst, and in a sense stabilizing criticality in an interval of coupling parameter.

Note that in the classical paradigm of Self-Organized Criticality the self-organization is reached at the price of adding hidden symmetries in the dynamics.  For example, in the discrete state sandpile model in \cite{bak-tang-etal:87} or its continuous state version \cite{zhang:89}, a local conservation law is imposed so as to reach, in the thermodynamic limit, the time scale separation between injection and dissipation necessary to achieve a critical state \cite{vespignani-zapperi:98,cessac-blanchard-etal:04}. Violating this conservation law with a local dissipation parameter $\epsilon$, breaks down criticality. But, there is a whole interval of $\epsilon$ where it is not possible to detect this violation in numerical, thus finite-sized, samples \cite{cessac-meunier:02}. By analogy, this would suggest that in our model, there might exist a local conservation law at the optimal noise, although we have not been able to assess what this law could be.

Yet, our work and its relations with critical systems needs further developments, subject of a forthcoming paper.
Especially, it is not enough to call "critical" a system exhibiting power laws. 
In statistical physics, critical systems are characterized by several critical exponents, which are related by relations obtained from the renormalization group analysis \cite{sethna:95,ma:01}. This imposes more complicated constraints than power-law emergence, leading to theoretical "crash tests", used on models to check whether they really exhibit a critical behaviour \cite{fontenele-vasconcelos-etal:19}. In the literature many models, sticking merely at the existence of power laws to prove criticality, are found not to pass these more elaborate tests \cite{touboul-destexhe:10b,cocchi-gollo-etal:17}. We believe that the mechanism exhibited here can be combined with an equation of transport for waves and a renormalization group analysis, in the spirit of  \cite{volchenkov-blanchard-etal:02}, to obtain constitutive relations between critical exponents.

\subsection{Two dimensions} \label{Sec:2D}
\begin{figure}
\centerline{
\resizebox{0.50\textwidth}{!}{
\includegraphics[height=6cm,width=7cm]{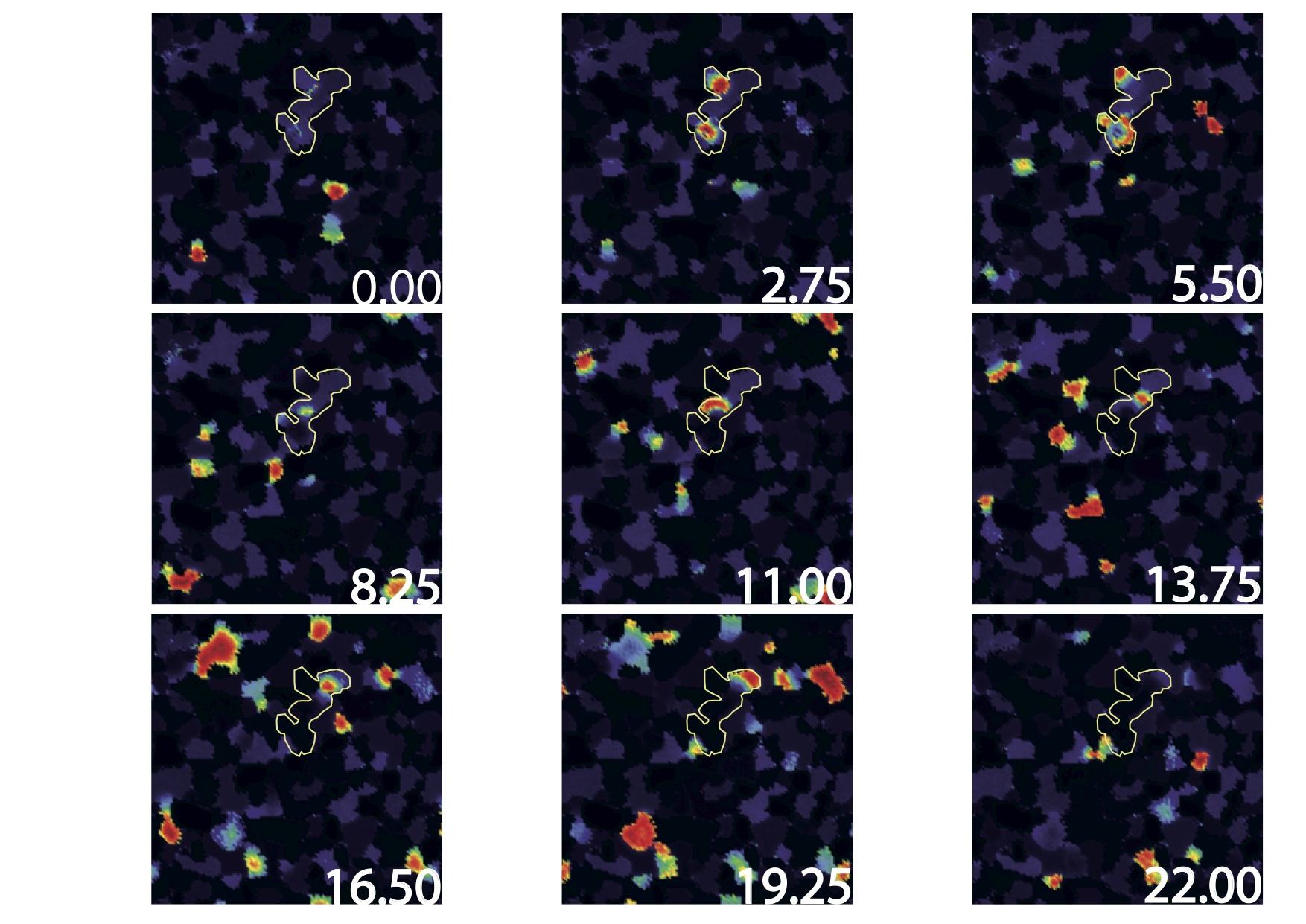}
}}
\caption{Example of the two dimensional time evolution of the calcium concentration $C$. Dark regions correspond to low calcium concentration while red corresponds to high concentration (wave). The time (in $s$) is displayed in the bottom right corner. The same thin white line in the center of each image, delimits a closed domain where wave propagates almost periodically. Hence, after the sequence shown above which correspond to a single period, a new one takes place a time later with a new wave following almost the same trajectory. The domain delimited by the white line is circled with high sAHP regions and therefore slowly evolves with time.}
\label{Fig:Movie}
\end{figure}

In section \ref{Sec:NumericalSimulations}, we already discussed why we limit ourselves to 1D numerical simulations. We want now to discuss the possible extensions of our results to genuine 2D type II retinal waves. In our model, a cell in the hyperpolarized state cannot be excited by its bursting neighbours. We observed in particular that the slow hyperpolarization current (sAHP) following the bursting phase, generates a landscape in which subsequent waves have to propagate. But the 1D and 2D cases are very different topologically. In 1D, regions of hyperpolarized cells with strong sAHP (high values of the variable $R$) constitute impassable borders to the propagation of the action potential. This is not the case in 2D, where they can be bypassed and give rise to more complex landscape. An example is shown in Fig. \ref{Fig:Movie}. 

In 1D, we have proved the existence of 2 thresholds for cholinergic coupling $g_{A}$, one associated with the stability of the travelling wave, the other with the nucleation of the travelling wave. In 2D, the travelling wave propagation speed is expected to depend on the local curvature of the front \cite{keener:86}. Only in the limit of a flat front we can hope to recover the results about the stability of an 1D travelling wave in absence of noise. However, it is not certain because in 2D, a flat front can generically destabilize itself through perturbations perpendicular to the front velocity (see \cite{saarloos:03} for a review). 

About the nucleation threshold, one may imagine that a sudden activity burst, due to a local synchronisation of SACs, gives rise to a centrifugal circular waves. As in 1D, the nucleation mechanism requires the synchronisation of high enough number of SACs firing together, but it also requests to overcome the front curvature effect, which can be a strong geometrical effect especially in the start-up phase. For all these reasons, we cannot analytically derive the 2D regime of parameters corresponding to the synchronization threshold from the 1D case. Nevertheless we succeed in numerically find a parameter regime where the avalanches size probability density function looks approaches a power law. A movie is available on line \cite{movie}. 

Furthermore, in the 1D case, we have found that, for the optimal noise amplitude ${ \widetilde \eta}$ and at the coupling strength threshold $g_{A}^{n}({\widetilde \eta})$, the avalanches size pdf follows a $P_{b^{*}}(S)={{\zeta(b^{*})}\over{S^{b^{*}}}}$ power law, with $b^{*}\simeq 3.17$. 
If we make a rough approximation and assume that the 2D analog is a centrifugal circular avalanche, then the probability to observe a 2D wave with size $S$ is related to the probability to observe a 1D wave with size $S^{{1}\over{2}}$, and therefore should decrease like 
${{1}\over{S^{{b^{*}}\over{2}}}}$. The exponent ${{b^{*}}\over{2}} \simeq 1.59$ is then not too much far compared to the values experimentally reported
\cite{hennig-adams-etal:09}.

\subsection{Understanding the functional role of retinal waves}\label{Sec:FuncRole}

In a critical regime waves exhibit maximal
variability in their sizes and durations. 
It would be interesting to explore, at a modelling level, what could be the functional impact of this regime in the development of the retino-thalamico-cortical pathways, elucidating why early neural networks would choose to maintain their activity in critical states. 
Our next step would be to link the retinal waves model presented here to the cortical V1 model developed in \cite{souihel-chavane-etal:19} to investigate the role of synaptic plasticity under retinal wave stimulation to shape the cortical response during developement.
More particularly, in our 2D numerical simulations, we observe (Fig. \ref{Fig:Movie}) the apparition of sAHP patterns that spatially bound the propagating waves. Those localized bounding patches are found to persist more than a sAHP timescale, indicating that in the network scale, they introduce a type of spatial memory of previous activity. In other words, they create an heterogeneous landscape where waves could propagate, inducing an immediate effect first on waves characteristics and second on the selective synaptic plasticity of the network through this spatial bias. Such patterns, although they have never been experimentally observed during stage II phase, are still interesting. For example, late waves (stage III) appear to be spatially bounded and more localised than stage II waves \cite{maccione-hennig-etal:14}, suggesting that probably late stage II waves could prepare the landscape for the transition to more spatially confined activity. Hence, those localized activity patterns, could potentially have a link with how receptive fields are formed before vision becomes functional.

\section*{Acknowledgements}
D.M.K. was supported by a doctoral fellowship from Ecole Doctorale des  Sciences et Technologies de l'Information et de la Communication de Nice-Sophia-Antipolis (EDSTIC).  This work also benefited from the support of the Neuromod institute of the University Côte d'Azur and the Doeblin federation. We warmly acknowledge Matthias Hennig,  Evelyne Sernagor, Olivier Marre, Serge Picaud for their invaluable help. 

\vfill\eject
\section{Appendix}
\subsection{Parameter Values}
\label{AppendixParameter}
\begin{table}[!h]
\begin{center}
    \begin{tabular}{ | p{2cm} | p{2.5cm} | }       			   \hline
    Parameter 	& 	Physical value 					\\ \hline
    $C_m$		& 	$22\, pF$  					\\ \hline
    $g_L$ 		& 	$2\, nS$ 	 					\\ \hline
    $g_{C}$	& 	$12 \, nS$						\\ \hline
    $g_{K}$		& 	$10 \, nS$						\\ \hline   
    $g_{H}$	& 	$\left[2,12\right] \, nS$ 			\\ \hline    
    $V_L$		& 	$-72 \, mV$					\\ \hline
    $V_{C}$	& 	$50 \, mV$ 					\\ \hline
    $V_K$		& 	$-90  \, mV$   					\\ \hline
    $V_1$		& 	$-20  \, mV$  					\\ \hline
    $V_2$		& 	$20 \, mV$ 					\\ \hline
    $V_3$		& 	$-25 \, mV$					\\ \hline
    $V_4$		& 	$7 \, mV$ 						\\ \hline
    $\tau_{N}$	&	$5 \, ms$    					\\ \hline
    $\tau_{R}$	&	$8250 \, ms$ 					\\ \hline
    $\tau_{S}$	&	$8250 \, ms$  					\\ \hline
    $\tau_{C}$	&	$2000 \, ms$  					\\ \hline
    $\delta_{C}$	&	$10.503 $ $nM \,\, pA^{-1}$ 		\\ \hline
    $\alpha_{S}$	&	$\frac{1}{200^4}$ $nM^{-4}$  		\\ \hline
    $\alpha_{C}$&	$4865$\, nM  					\\ \hline
    $\alpha_{R}$&	$4.25$  						\\ \hline
    $H_X$		&	$1800$ $nM$					\\ \hline
    $C_0$		&	$88$ $nM$					 \\ \hline
    $\eta$	    &	$[5.2,9.0] \,pA ms^{1/2}$	     \\ \hline
    \end{tabular}
\end{center}
\caption{Typical parameters values used in eq.(\ref{eq:FullModel}).}
\label{TabParameters}
\end{table}

\begin{table}[!h]
\begin{center}
    \begin{tabular}{ | p{2cm} | p{2.5cm} | }       			   \hline
    Parameter 		& 	Physical value 					\\ \hline
    $\mu$			& 	$ 1.82 \,s$  					\\ \hline
    $K_{Ach}$		&	$200 \, V^{-1}$					\\ \hline
    $V_{A}$		& 	$0 \, mV$						\\ \hline
    $V_{0}$			& 	$-40 \, mV$					\\ \hline
    $\gamma_{Ach}$ 	& 	$1 \,nM^2$					\\ \hline
    $\beta_{Ach}$ 	& 	$5 \,nM \,s^{-1}$				\\ \hline
    \end{tabular}
\end{center}
\caption{Typical parameters values related to the acetylcholine neurotransmitter (Ach).}
\label{TabParametersNext}
\end{table}

\subsection{Wave threshold}
\label{AppendixThreshold}
Denoting $V_-$ the rest state of a SAC, the condition for cell $j$ to start bursting at time $t_B$, in the absence of noise,  is $I_{tot,j}(t_B)=I_{SN}$ which reads:
\begin{equation}\label{eq:CondStartBurst}
\begin{array}{lll}
&-g_{A} \left(V_--V_{A}\right) \displaystyle{\sum_{k \in \left\{ j \right\} } {{A_{k}(t_B)^{2}}\over{\gamma_{Ach}+A_{k}(t_B)^2}}}\\
& -g_{H} R_j^4(t_B) \pare{V_--V_K}
 = I_{SN}.
 \end{array}
\end{equation}

This equation nicely illustrates the competitive role of the excitatory Ach current coming from neighbouring bursting SACs and the slow hyperpolarization current. The more a cell is hyperpolarized the harder it is to excite it by Ach and there is a long hyperpolarization phase during which a SAC cannot be excited by its excited neighbours. The wave of activity propagates therefore into a landscape of sAHP coming from previous waves, rendering the dynamics dependent on the network history on large time scales (several minutes). Additionally, when two waves collide, they penetrate each other until they reach the hyperpolarization zone, where they stop. 

Eq. \eqref{eq:CondStartBurst} allows to define a critical value for $g_{A}$ below which it is not possible to propagate a wave. In the absence of acetylcholine current the rest state $V_-$ obeys the implicit non-linear equation: 
\begin{equation}\label{eq:RestStateV}
V_- =
\frac{ g_{L} V_{L} + g_{C} M_- V_{C} +  V_{K} \pare{g_K N-  + g_{H}R_-^4}}{g_L+g_{C} M_- + g_{K} N_-  + g_{H}R_-^4},
\end{equation}
with
 \begin{equation}\label{eq:FixedPoint}
\left\{
\begin{array}{lll}
M_- &=&M_\infty\bra{V_-} \\
N_-&=&N_{\infty}(V_-)\\
C_-&=&\frac{H_X}{\alpha_C} \bra{C_0 - \delta_{C}  g_{C} M_-(V_--V_{C})}\\
S_- &=&\frac{\alpha_S C^{4}_-}{1+\alpha_S C^{4}_-}\\
R_-&=&\frac{\alpha_{R}S_-}{1+\alpha_{R} S_-}.\\
\end{array}
\right.
\end{equation}
Let $A_+$ be the maximal possible production of Ach and $n$ the number of neighbours. 
In the absence of noise, a propagation is possible only if:
\begin{equation}\label{eq:gAmin}
g_{A} \geq \frac{I_{SN}+g_{H}R_-^4 \pare{V_--V_K}}{\frac{n \, A_+^2}{\gamma_{Ach}+A_+^2} \, \pare{V_{A}-V_-}}.
\end{equation}
In this equation, one assumes that all neighbors are active with a maximal Ach production. This is therefore a lower threshold. 
In the presence of noise this threshold is not sharp anymore: the probability to trigger an avalanche is a sigmoid whose slope increases as $\eta \to 0$.
\end{document}